# Level-crossing induced spin phenomena in SiC: a theoretical study

Denis V. Sosnovsky, Konstantin L. Ivanov*

*International Tomography Center, Siberian Branch of the Russian Academy of Sciences, Novosibirsk, 630090, Russia*

*Novosibirsk State University, Novosibirsk, 630090, Russia*

* Corresponding author, email: ivanov@tomo.nsc.ru

**Abstract**
A theoretical approach is proposed to treat the spin dynamics in defect color centers. The method explicitly takes into account the spin dynamics in the ground state and excited state of the defect center as well as spin state dependent transitions involving the ground state and excited state, as well as an additional intermediate state. The proposed theory is applied to treat spin-dependent phenomena is silicon carbide, namely, in spin-$\frac{3}{2}$ silicon-vacancy centers, $V_{Si}$ or V2 centers. Theoretical predictions of magnetic field dependent photoluminescence intensity and optically detected magnetic resonance spectra demonstrate an important role of level crossing phenomena in the spin dynamics of the ground state and excited state. The results are in good agreement with previously published experimental data [Phys. Rev. X, 6 (2016) 031014].

# I. Introduction

Defect color centers are promising objects for various applications, such as optical detection of magnetic resonance [1-3], quantum information processing [4,5], nanosensing [6-11] and nuclear spin hyperpolarization [12-17]. An example of a particularly well studied and widely used system of this kind is given by the negatively charged nitrogen-vacancy centers, NV⁻ centers, in diamond crystals [1,2]. However, defect color centers in other crystal systems also hold a great promise for similar applications, notably, defect centers in silicon carbide (SiC) and boron nitride. Previous works on such systems suggest that the color centers in these crystals can be utilized to perform Optically Detected Magnetic Resonance (ODMR) experiments and to probe nano-scale environment.

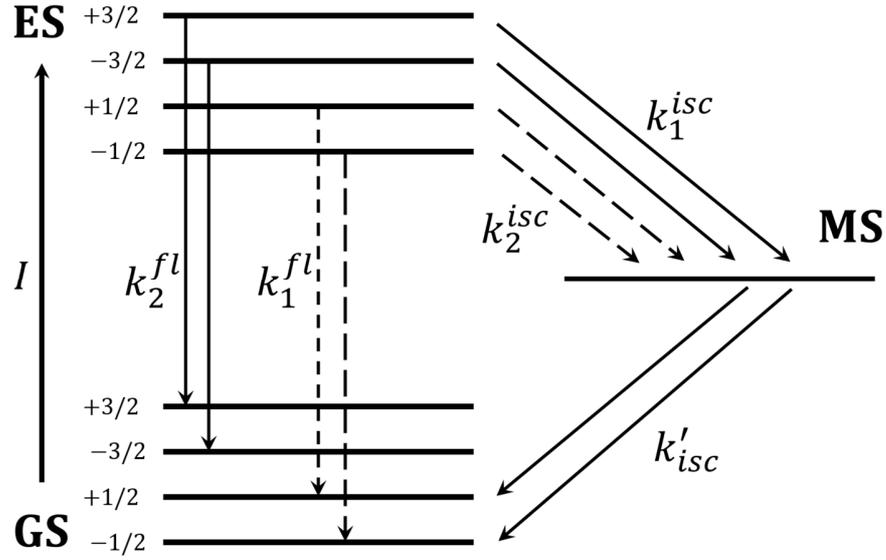

*Scheme 1. Energy level diagram of the silicon-vacancy center in SiC, having the $S=\frac{3}{2}$ electronic ground state and excited state. Transitions between these two states occur at a rate $I$ due to photo-excitation; transitions from the excited state to the ground state are also possible via fluorescence, with the fluorescence rate depending on the spin state of the defect center, i.e., $k_1^{fl} \neq k_2^{fl}$. Transitions to the ground state are also possible via another route: ISC from the excited state to an intermediate state MS with $k_1^{isc} \neq k_2^{isc}$ followed by ISC from this state to the ground state at a rate $k'_{isc}$. Due to the spin-selective character of the transitions, after a few excitation cycles the defect center becomes spin-polarized.*

This work is focused on spin phenomena in SiC [18], specifically, we study silicon-vacancy centers, V$_{Si}$ or V2 centers [11,19-22]. These are negatively charged paramagnetic spin-$\frac{3}{2}$ defect states; according to electron-nuclear double resonance experiments [23], they are non-covalently bonded to diamagnetic neutral carbon vacancies located at the adjacent site along the SiC c-axis. Such defect centers are promising candidates for various applications [24-30], including coherent control of optically addressable spin qubits at room temperature, magnetometry and thermometry. The aim of this work is to develop a theoretical framework for describing spin phenomena in such system, notably, phenomena originating from spin mixing at level crossings. The complexity of the problem under study comes from two factors: (i) interplay between the spin dynamics in the excited and ground state of the color center; (ii) the presence of high-spin states, specifically, of $S=\frac{3}{2}$ states. To solve this problem, we generalize an approach [31], which we used earlier to describe optically induced spin polarization formed in NV⁻ centers and in triplet states in molecular crystals. In this method we explicitly treat the spin dynamics in the ground state and excited states; the transitions between the ground and excited states are introduced using the Lindblad formalism [32,33], which can be efficiently exploited to describe relaxation phenomena in color centers [34-36]. To gain insight into the spin dynamics, we also look at Level Crossings (LCs) of the spin states and determine under what conditions they are turned into Level Anti-Crossings (LACs). As discussed previously, LACs give rise to pronounced features in the magnetic field dependence of photoluminescence (PL) of color centers [37-43] and in ODMR spectra [43]. We demonstrate that the proposed theory can be used to model spin-dependent phenomena in defect centers in SiC, giving a good agreement between calculated and experimental [43] ODMR spectra and magnetic field dependent PL intensity.

## II. Theory

### A. Theoretical framework

In this work, we consider the energy level diagram of the V$_{Si}$ color center, shown in **Scheme 1**. The system undergoes transitions between the ground state, GS, and excited state, ES, due to light-excitation (inducing reversible transitions occurring at a rate $I$); from ES the system goes back to GS via radiative transitions, giving rise to fluorescence. The rate of these transitions depends on the spin state of the defect center, being equal to $k_1^{fl}$ and $k_2^{fl}$ for the $S_z = \pm\frac{1}{2}$ states and $S_z = \pm\frac{3}{2}$ states, respectively. The transition from ES→GS can take another route: Inter-System Crossing (ISC) to an intermediate state MS followed by ISC to the ground state. We also assume that the rate of the ISC process from ES to the intermediate state differs for the $S_z = \pm\frac{1}{2}$ states and $S_z = \pm\frac{3}{2}$ states (the corresponding rates are denoted as $k_1^{isc}$ and $k_2^{isc}$, respectively) and that from the intermediate state transitions are possible only to the $S_z = \pm\frac{1}{2}$ spin states in GS (with the rate denoted $k'_{isc}$). These assumptions are consistent with literature data [21]. Due to the difference in the rates, namely $k_1^{fl} \neq k_2^{fl}$, $k_1^{isc} \neq k_2^{isc}$ and $k'_{isc} \neq 0$ only for the $S_z = \pm\frac{1}{2}$ states, after several excitation cycles the V$_{Si}$ defect acquires spin polarization [21], that is, the populations of the $S_z = \pm\frac{3}{2}, \pm\frac{1}{2}$ states become different in GS and in ES. For the sake of generality, we also assume that there is a defect center also has a magnetic spin-$\frac{1}{2}$ nucleus coupled to the unpaired electrons.

In all cases, we consider the situation where the symmetry axis of the defect center is parallel (or nearly parallel) to the external magnetic field, which is set parallel to the $Z$-axis of the lab frame. As a consequence, the $S_z = \pm\frac{3}{2}$ states and $S_z = \pm\frac{1}{2}$ states are eigenstates of the ES and GS Hamiltonians not only at zero field, but at all magnetic fields, except for the ones corresponding to LACs (see discussion below). At LACs, spin mixing between the $S_z = \pm\frac{3}{2}$ and $S_z = \pm\frac{1}{2}$ states occurs.

For modeling spin-dependent phenomena in V$_{Si}$ centers we used an approach described in our previous paper [31]. First, we introduce the density matrix and Hamiltonian of the defect center of a block-diagonal form:

$$\rho_{SiC} = \begin{pmatrix} \rho_{GS} & 0 & 0 \\ 0 & \rho_{ES} & 0 \\ 0 & 0 & \rho_{MS} \end{pmatrix}, \quad \widehat{\mathcal{H}}_{SiC} = \begin{pmatrix} \widehat{\mathcal{H}}_{GS} & 0 & 0 \\ 0 & \widehat{\mathcal{H}}_{ES} + \mathcal{E}_{ES}\hat{1} & 0 \\ 0 & 0 & \widehat{\mathcal{H}}_{MS} + \mathcal{E}_{MS}\hat{1} \end{pmatrix} \quad (1)$$

Hence, we introduce the density matrix and the Hamiltonian in the basis of spin states $|\mu_{GS}\rangle, |\mu_{ES}\rangle, |\mu_{MS}\rangle$ defined for each electronic state GS, ES and MS. If we list these states, the basis becomes $|1_{GS}\rangle, ..., |N_{GS}\rangle, |1_{ES}\rangle, ..., |N_{ES}\rangle, |1_{MS}\rangle, ..., |N'_{MS}\rangle$. Here $N_i$ is the number of spin states in the corresponding state, which is equal to $N$ for GS and ES and to $N'$ for MS ($N' \neq N$ because of the different multiplicity of the electronic states). The form of the density matrix implies that off-diagonal elements of the density matrix can be non-zero for the spin states belonging to the same state GS, ES or MS, while there are no coherences between the spin

states belonging ES, GS and MS (these matrix elements are subject to very fast decoherence). Hence, for each state (GS, ES and MS) we introduce its spin density matrix ($\rho_{GS}$, $\rho_{ES}$ and $\rho_{MS}$, respectively). For the sake of simplicity, we also neglect completely the spin dynamics in the MS state. In this situation, we can omit the actual structure of the MS quantum states and set $N' = 2$, corresponding to the two states of the spin-$\frac{1}{2}$ nucleus. This assumption is justified because in MS there is no LAC-driven spin dynamics.

In the Hamiltonian matrix, see eq. (1), we introduce the energy splitting between GS and ES as $\mathcal{E}_{ES}$ and the energy splitting between GS and MS is denoted as $\mathcal{E}_{MS}$. The spin dynamics in each state is described by the corresponding spin Hamiltonian $\widehat{\mathcal{H}}_{GS}$, $\widehat{\mathcal{H}}_{ES}$ and $\widehat{\mathcal{H}}_{MS}$; hereafter $\hat{1}$ is the unity matrix of the corresponding dimensionality. In the numerical solution procedure, the energies $\mathcal{E}_{ES}$ and $\mathcal{E}_{MS}$ can be omitted (since we are not interested in the coherences between GS, ES and MS and their evolution).

The density matrices $\rho_{GS}$ and $\rho_{ES}$, see eq. (1), are introduced in the state basis, which is the direct product of the quartet basis $|Q\rangle = \left\{\left|\frac{3}{2}\right\rangle, \left|\frac{1}{2}\right\rangle, \left|-\frac{1}{2}\right\rangle, \left|-\frac{3}{2}\right\rangle\right\}$ for the electron spin states and the Zeeman basis for the nuclear spin states $|Z\rangle$; in the simplest case of a single spin-$\frac{1}{2}$ nucleus $|Z\rangle = \left\{\left|\frac{1}{2}\right\rangle, \left|-\frac{1}{2}\right\rangle\right\} = \{|\alpha\rangle, |\beta\rangle\}$ (here $|\alpha\rangle$ and $|\beta\rangle$ are the standard notations for the "spin-up" and "spin-down" states). For $\rho_{MS}$ the basis is introduced as the nuclear Zeeman basis. The spin Hamiltonians describe the Zeeman interaction of spins with the external field $\mathbf{B}$, Zero-Field Splitting (ZFS) of the spin-$\frac{3}{2}$ states and electron-nuclear Hyper-Fine Coupling (HFS) with the magnetic nucleus of the color center (extension to $K$ magnetic nuclei is straightforward). In the case $\mathbf{B}||Z$ the Hamiltonians of the quartet states are as follows, when written in $\hbar$ units (here we keep in mind that the $E$-parameters in the ZFS tensor are zero, owing to the $C_{3v}$ symmetry of the system):

$$\widehat{\mathcal{H}}_{GS} = \gamma_e B \hat{S}_z - \gamma_N B \hat{I}_z + D_G \left\{\hat{S}_z^2 - \frac{5}{4}\right\} + \hat{\mathbf{S}} \widehat{\mathcal{A}}^{(GS)} \hat{\mathbf{I}} \qquad (2)$$
$$\widehat{\mathcal{H}}_{ES} = \gamma_e B \hat{S}_z - \gamma_N B \hat{I}_z + D_E \left\{\hat{S}_z^2 - \frac{5}{4}\right\} + \hat{\mathbf{S}} \widehat{\mathcal{A}}^{(ES)} \hat{\mathbf{I}}$$

Here $\hat{\mathbf{S}}$ is the electron spin operator of the spin-$\frac{3}{2}$ state and $\hat{\mathbf{I}}$ is the spin operator of the magnetic nucleus. The first two terms stand for the electron and nuclear Zeeman interactions with $\gamma_e = g_{\parallel}^{(1)} \mu_B$ and $\gamma_N = g_N \mu_N$ being the corresponding gyromagnetic ratios; here $\mu_B$ and $\mu_N$ stand for the Bohr magneton and nuclear magneton, respectively; $g_{\parallel}^{(1)}$ and $g_N$ are the corresponding $g$-factors (the subscript || stands for the $g$-value for the orientation of the defect symmetry axis along the magnetic field). The D-terms stand for the ZFS, the last terms stand for HFSs with $\widehat{\mathcal{A}}^{(GS/ES)}$ being the HFS tensor in GS and ES. The $D$-parameter for the two GS and ES states is equal to $D_G = 1.25$ mT and $D_E = 7.32$ mT, respectively [21]. The eigen-states of the ZFS Hamiltonian are the $\left|\frac{3}{2}\right\rangle, \left|\frac{1}{2}\right\rangle, \left|-\frac{1}{2}\right\rangle, \left|-\frac{3}{2}\right\rangle$ states for both GS and ES. One should also note that the HFC-tensors are generally different for the two triplet states, being equal to $\widehat{\mathcal{A}}^{(GS)}$ and $\widehat{\mathcal{A}}^{(ES)}$. In some cases, we also add various perturbation terms to the Hamiltonians, $\hat{V}_{GS}$ and $\hat{V}_{ES}$:

$$\hat{V}_{GS/ES} = g_{\parallel}^{(2)} \mu_B \hat{S}_z \left(\hat{S}_z^2 - \frac{5}{4}\right) B + g_{\parallel}^{(3)} \mu_B \frac{\hat{S}_+^3 - \hat{S}_-^3}{4i} B + g_{\perp}^{(1)} \mu_B \hat{\mathbf{S}}_\perp \hat{\mathbf{B}}_\perp + g_{\perp}^{(2)} \mu_B \left\{\hat{\mathbf{S}}_\perp \hat{\mathbf{B}}_\perp, \left(\hat{S}_z^2 - \frac{3}{4}\right)\right\} \qquad (3)$$
$$- \frac{i}{4} g_{\perp}^{(3)} \mu_B \left(\{\hat{S}_+^2, \hat{S}_z\} B_+ - \{\hat{S}_-^2, \hat{S}_z\} B_-\right)$$

Here $\hat{\mathbf{S}}_\perp = (\hat{S}_x, \hat{S}_y)$, $\mathbf{B}_\perp = (B_x, B_y)$, $\hat{S}_\pm = \hat{S}_x \pm i\hat{S}_y$, $B_\pm = B_x \pm iB_y$, $\{\hat{A}, \hat{B}\} = \hat{A}\hat{B} + \hat{B}\hat{A}$ stands for the anti-commutator of two operators. The six $g$-factors introduced above are linearly independent in systems of the $C_{3v}$ symmetry [21]. Typically, $g_{\parallel}^{(1)}, g_{\perp}^{(1)} \approx 2$, while the other four $g$-factors are much smaller than unity [21].

The temporal evolution of density matrix of the entire system is described by the Liouville-von Neumann equation

$$\frac{d\rho_{SiC}}{dt} = -i[\hat{\mathcal{H}}_{SiC}, \rho_{SiC}] + \hat{\hat{\mathcal{R}}}\rho_{SiC} \quad (4)$$

The super-operator $\hat{\hat{\mathcal{R}}}$ describes the transitions between different states, here GS, ES and MS, due to light-excitation, luminescence and ISC. It can be written in the Lindblad form, which comprises two terms:

$$\hat{\hat{\mathcal{R}}}\rho_{SiC} = \{\hat{\hat{\mathcal{R}}}_1 + \hat{\hat{\mathcal{R}}}_2\}\rho_{SiC}, \quad (5)$$

$$\hat{\hat{\mathcal{R}}}_1\rho_{SiC} = -\frac{1}{2}\sum_{m,n}\{\hat{\mathcal{L}}_{mn}^\dagger \hat{\mathcal{L}}_{mn}\rho_{SiC} + \rho_{SiC}\hat{\mathcal{L}}_{mn}^\dagger \hat{\mathcal{L}}_{mn}\}, \quad \hat{\hat{\mathcal{R}}}_2\rho_{SiC} = \sum_{m,n}\hat{\mathcal{L}}_{mn}\rho_{SiC}\hat{\mathcal{L}}_{mn}^\dagger$$

These two terms stand for escape from a given state and income to another state. The operators $\hat{L}_{mn}$ are defined by introducing the rates, $k_{mn}$, of transitions between the quantum states $|m\rangle \to |n\rangle$. They have the following non-zero elements:

$$\langle n|\hat{L}_{mn}|m\rangle = \sqrt{k_{mn}} \quad (6)$$

The relevant rates of the processes are indicated in **Scheme 1**. We explain in detail now to introduce the $\hat{L}_{mn}$ operators in **Supplementary Materials**.

For calculating the PL intensity, we first evaluate the steady-state value of the density matrix $\rho_{SiC}$. To do so, we solve eq. (4) setting $\frac{d}{dt}\rho_{SiC} = 0$. To obtain a solution, which is different from the trivial solution $\rho_{SiC} = 0$, we act as follows. Since the rank of the $\left(-i\hat{\hat{\mathcal{H}}}_{SiC} + \hat{\hat{\mathcal{R}}}\right)$ super-matrix is equal to $(M^2 - 1)$ (when $\rho_{SiC}$ is an $M \times M$ matrix), one of the equations in the system $\frac{d}{dt}\rho_{SiC} = 0$ is linearly dependent on other equations. Therefore, in order to obtain the solution for $\rho_{SiC}$ we replace the last equation in this system by expression $\sum_i\{\rho_{SiC}\}_{ii} = \text{Tr}\{\rho_{SiC}\} = 1$, which provides normalization of the density matrix. The new system of linear equations is straightforward to solve and $\rho_{SiC}$ can be obtained [31]. Knowing $\rho_{SiC}$ we can compute all experimental observables of interest. For instance, the photo-luminescence intensity is given by the following expression

$$I_{PL} = k_1^{fl}\text{Tr}\{\hat{P}_1^{ES}\rho_{SiC}\} + k_2^{fl}\text{Tr}\{\hat{P}_2^{ES}\rho_{SiC}\} \quad (7)$$

That is, we multiply the luminescence rate from a specific state by the population of this state. Here $\hat{P}_1^{ES}$ and $\hat{P}_2^{ES}$ are the projector operators onto the $\left|\pm\frac{1}{2}\right\rangle$ and $\left|\pm\frac{3}{2}\right\rangle$ spin states in ES, respectively. In experiments one can also measure the variation of the PL intensity, applying a weak oscillating magnetic field $\Delta B(t) = \delta B \cdot \cos(\omega_{mod}t)$ field along the $Z$-axis and using lock-in detection at the $\omega_{mod}$ frequency. In this case, instead of the dependence $I_{PL}(B)$ one would measure the field dependence of $\frac{d}{dB}I_{PL}$, i.e., of the derivative of $I_{PL}$.

To calculate the ODMR signal we proceed as follows. First, we add terms describing the interaction with the circularly polarized transverse radiofrequency (RF) field, the $B_1$-field:

$$\hat{\mathcal{H}}_{GS} \to \hat{\mathcal{H}}_{GS} + \gamma_e B_1\left(\cos[\omega_{rf}t]\hat{S}_x + \sin[\omega_{rf}t]\hat{S}_y\right) \quad (8)$$

$$\hat{\mathcal{H}}_{ES} \to \hat{\mathcal{H}}_{ES} + \gamma_e B_1\left(\cos[\omega_{rf}t]\hat{S}_x + \sin[\omega_{rf}t]\hat{S}_y\right)$$

Subsequently, we go to the rotating frame, in which the Hamiltonians are written as follows:

$$\hat{\mathcal{H}}'_{GS} = (\gamma_e B - \omega_{rf})\hat{S}_z - \gamma_e B_1 \hat{S}_x - \gamma_N B \hat{I}_z + D_G\left\{\hat{S}_z^2 - \frac{5}{4}\right\} + \hat{\mathbf{S}}\hat{\mathcal{A}}_{sec}^{(GS)}\hat{\mathbf{I}} \qquad (9)$$

$$\hat{\mathcal{H}}'_{ES} = (\gamma_e B - \omega_{rf})\hat{S}_z - \gamma_e B_1 \hat{S}_x - \gamma_N B \hat{I}_z + D_E\left\{\hat{S}_z^2 - \frac{5}{4}\right\} + \hat{\mathbf{S}}\hat{\mathcal{A}}_{sec}^{(ES)}\hat{\mathbf{I}}$$

Here we subtract $\omega_{rf}\hat{S}_z$ from the electronic Zeeman term; in the HFC terms we keep only their secular parts containing the $\hat{S}_z$ operator: the terms containing the $\hat{S}_{x,y}$ operators in the rotating frame are multiplied by fast oscillating exponents $e^{\pm i\omega_{rf}t}$ and can be omitted. It is important to note that transformation to the rotating frame is used only for the electron spins. Using the Hamiltonians from eq. (9) we calculate the steady-state value of $\rho_{SiC}$ and the PL intensity. The ODMR signal is then proportional to the variation of PL intensity:

$$ODMR \sim I_{PL}(rf/on) - I_{PL}(rf/off) \qquad (10)$$

where $I_{PL}(rf/on)$ and $I_{PL}(rf/off)$ is the PL intensity measured in the presence $B_1 \neq 0$ and in the absence $B_1 = 0$ of the RF-field. By plotting the ODMR signal as a function of $\omega_{rf}$ we obtain the ODMR spectrum. When the energy separation of the corresponding levels coincides with the frequency of the applied RF-field, a spin resonance transition can occur by either absorbing or emitting an RF photon. These RF-induced spin transitions can now be detected by monitoring variation of the total PL intensity.

*Table 1. Parameters of the $V_{Si}$ center used in calculations.*

| Reaction rates [ns$^{-1}$] | $I = 0.01$, $k'_{isc} = 0.01$, $k_2^{fl} = 2k_1^{fl} = 0.1$, $k_1^{isc} = 20k_2^{isc} = 0.2$ |
|---|---|
| ZFS parameters [mT] | $D_G = 1.25$, $D_E = 7.32$ |
| HFC tensor of $^{29}$Si [mT] | $A_{xx}^{(GS)} = A_{yy}^{(GS)} = A_{zz}^{(GS)} = 0.001$; $\quad A_{xx}^{(ES)} = A_{yy}^{(ES)} = A_{zz}^{(ES)} = 0.001$ |

In some cases, for sake of simplicity, we evaluate the ODMR signal not from eq. (10); instead we evaluate the difference of the populations $\delta P_{ij}$ of the relevant states $|i\rangle$ and $|j\rangle$ involved in the resonant transition. Assuming that the RF-field only slightly perturbs the populations, we calculate the relative ODMR signal, which is given by the intensity ratio $\frac{S_1}{S_2}$, for the transitions $\left|-\frac{3}{2}\right\rangle \leftrightarrow \left|-\frac{1}{2}\right\rangle$ and $\left|\frac{3}{2}\right\rangle \leftrightarrow \left|\frac{1}{2}\right\rangle$ in the ground state. In turn, the intensities of these transitions are evaluated as the population difference for the corresponding states:

$$S_1 = \text{Tr}\left\{\hat{P}_{-3/2}^{GS}\rho_{SiC} - \hat{P}_{-1/2}^{GS}\rho_{SiC}\right\}, \qquad S_2 = \text{Tr}\left\{\hat{P}_{3/2}^{GS}\rho_{SiC} - \hat{P}_{1/2}^{GS}\rho_{SiC}\right\} \qquad (11)$$

Here $\hat{P}_{\pm 1/2}^{gr}$ and $\hat{P}_{\pm 3/2}^{gr}$ are the projector operators on the $\left|\pm\frac{1}{2}\right\rangle$ and $\left|\pm\frac{3}{2}\right\rangle$ ground electronic states, respectively.

Finally in this subsection, we introduce the typical parameters used in the calculation. The relevant rates, ZFS parameters and HFC parameters are listed in **Table 1**. For the HFC we assume that the paramagnetic defect center is coupled to a single $^{29}$Si nucleus (spin-$\frac{1}{2}$ nucleus) of the lattice. The HFC tensor is taken symmetric in most cases and the HFC constant is taken the same in GS and ES. Unless otherwise stated, we use the parameters from **Table 1**; when this is not the case, we add a clarifying note. The ZFS parameters have been determined in previous works, while the rates follow from fitting experimental data, presented in **Section III**. The reaction rates were chosen to give the best agreement with the experimental data. It is important to note that for different relation between the rates $k_1^{fl}$ and $k_2^{fl}$ and also $k_1^{isc}$ and $k_2^{isc}$ the field dependence of the PL intensity and the shape of the ODMR spectrum vary substantially: LAC-derived features come up as peaks or dips, depending on the precise values of the rates. The parameters given in **Table 1** provide the proper appearance of all these features, which is consistent with available experimental data [43]. The rates that we use are in reasonable agreement with previously reported values [44]. In **Supplementary Materials** we provide detailed comments on the choice of the relevant parameters.

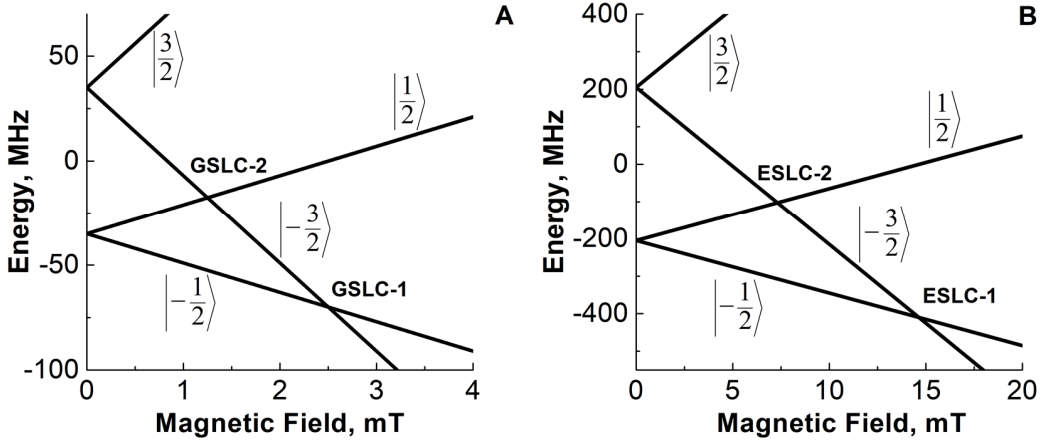

*Figure 1. Energy levels and LCs in the $V_{Si}$ center shown for GS (A) and ES (B).*

**B. LAC analysis**

To gain understanding of the spin phenomena, it is useful to support numerical calculations by the analysis of level crossings. When the defect center is oriented parallel to the external field, there are LCs in both GS and ES, occurring at well-defined magnetic fields, see **Figure 1**. In each case, there are two LCs, which we name GSLCs and ESLCs. Additional terms would turn these LCs into LACs, as discussed below.

To obtain the LC field positions, we proceed as follows. We omit all terms in the Hamiltonian, except for the electron Zeeman interaction and ZFS interaction. In this situation, the eigenproblem of the Hamiltonian can be solved analytically; at all field the eigenstates are the states with a specific $S_z$ value, which is $\pm\frac{1}{2}$ or $\pm\frac{3}{2}$. From the solution, one can obtain [21] that at $D > 0$ there is a crossing between the levels corresponding to the $\left|-\frac{3}{2}\right\rangle$ and $\left|-\frac{1}{2}\right\rangle$ states, which occurs when $\omega_e = \gamma_e B = 2D$ ($\Delta S_z = 1$, hereafter LC-1), and to the $\left|-\frac{3}{2}\right\rangle$ and $\left|\frac{1}{2}\right\rangle$ states, which occurs when $\omega_e = \gamma_e B = D$ ($\Delta S_z = 2$, hereafter LC-2). In the case under study, we obtain four LC positions, see **Figure 1**, corresponding to GSLC-1 at $B = 2D_G/\gamma_e \approx 2.5$ mT, GSLC-2 at $B = D_G/\gamma_e \approx 1.25$ mT (see **Figure 1A**), to ESLC-1 at $B = 2D_E/\gamma_e \approx 14.64$ mT and to ESLC-2 at $B = D_E/\gamma_e \approx 7.32$ mT (see **Figure 1B**). If we introduce HFC with the spin-$\frac{1}{2}$ magnetic nucleus, each LC of the electron spin states will split into four LCs, corresponding to different nuclear spin states. In the present work, we consider only the case of small HFC, so that this splitting (which is not observed in any of the experiments reported up to date) is vanishingly small.

Various perturbations cause mixing of states and turn the LCs discussed above into LACs: the crossings are avoided due to the presence of the perturbation terms, which mix the states. The minimal splitting of the states at the LAC is equal to $2|V_{mn}|$, where $V_{mn} = \langle m|\hat{V}|n\rangle$ is the corresponding matrix element of $\hat{V}$ between the unperturbed states $|m\rangle$ and $|n\rangle$. The perturbation term becomes relevant when $V_{mn}$ is greater than or comparable to the splitting $\delta\mathcal{E}_{mn}$ between the unperturbed energy levels: this condition defines the width of the LAC region. What is important in the context of the present work, at the LAC the states $|m\rangle$ and $|n\rangle$ are no longer the eigenstates of the Hamiltonian. As a consequence, their populations become mixed. In the case of the $V_{Si}$ center, the crossing states are characterized by different $S_z$ values and have different populations due to the properties of the optical cycle of the color center, as discussed in **Scheme 1**. Excitation from the different states also gives rise to different PL intensity. Upon spin mixing at LACs, the populations are redistributed giving rise increase or reduction of the PL intensity. Hence, LACs are expected to give rise to features, such as peaks or dips, in the $I_{PL}(B)$ curve.

A detailed analysis of the role of different perturbations, which turn LCs into LACs, is presented in **Supplementary Materials**. In this analysis, we calculate the relevant $V_{mn}$ matrix elements for different perturbation terms and compare the results of the LAC analysis with numerical results.

Below we consider in detail a single case, which is relevant for the V$_{Si}$ centers. Let us assume that the $Z$-axis of the ZFS tensor is not perfectly aligned along the $Z$-axis of laboratory frame, i.e., with the external magnetic field. In this case, a perturbation term emerges, which can turn LCs into LACs. The perturbation term can be written as follows:

$$\hat{V}_{\perp}^{(1)} = \gamma_e B \theta \hat{S}_x \tag{12}$$

Here $\theta$ is the angle between the $Z$-axis of the ZFS tensor and the external magnetic field (we assume that the **B** vector lies in the $XZ$-plane); this angle is taken small, hence $\sin\theta \approx \theta$. One should note that the full Zeeman Hamiltonian comprises [21] additional terms

$$\hat{V}_{\parallel}^{(2)} = g_{\parallel}^{(2)} \mu_B \hat{S}_z \left(\hat{S}_z^2 - \frac{5}{4}\right) B \quad \text{and} \quad \hat{V}_{\perp}^{(2)} = g_{\perp}^{(2)} \mu_B B \theta \left\{\hat{S}_x, \left(\hat{S}_z^2 - \frac{3}{4}\right)\right\}$$

which emerge due to non-equivalence of the $Z$-axis and the perpendicular $X,Y$-axes, and terms

$$\hat{V}_{\parallel}^{(3)} = g_{\parallel}^{(3)} \mu_B \frac{\hat{S}_+^3 - \hat{S}_-^3}{4i} B \quad \text{and} \quad \hat{V}_{\perp}^{(3)} = -\frac{i}{4} g_{\perp}^{(3)} \mu_B B \theta \{(\hat{S}_+^2 - \hat{S}_-^2), \hat{S}_z\}$$

which are coming from the trigonal pyramid symmetry of the defect center. The term $\hat{V}_{\parallel}^{(2)}$ does not cause any mixing of the electronic spin states, moreover $g_{\parallel}^{(2)} \approx 0$. The term $\hat{V}_{\parallel}^{(3)}$ causes mixing of only the states $\left|\frac{3}{2}\right\rangle$ and $\left|-\frac{3}{2}\right\rangle$ and due to the small value of $g_{\parallel}^{(3)}$ (see **Supplementary Materials**) slightly shifts the position of the relevant LCs . We neglect this term for the sake of simplicity. The term $\hat{V}_{\perp}^{(2)}$ gives rise to the same effect as the perturbation $\hat{V}_{\perp}^{(1)}$. However, because of the small value of $g_{\perp}^{(2)}$ (see **Supplementary Materials**), the corresponding matrix element is negligible. The term $\hat{V}_{\perp}^{(3)}$ is small but nevertheless gives rise to the spin mixing in the first-order perturbation theory, between the states $\left|-\frac{3}{2}\right\rangle$ and $\left|\frac{1}{2}\right\rangle$. In calculation $g_{\perp}^{(3)}$ is taken equal to 0.2 (as explained in **Supplementary Materials**). Hence, the total misalignment term is given by $\hat{V}_{\perp} = \hat{V}_{\perp}^{(1)} + \hat{V}_{\perp}^{(2)} + \hat{V}_{\perp}^{(3)}$. The perturbation terms $\hat{V}_{\perp}^{(1)}$ and $\hat{V}_{\perp}^{(2)}$ cause mixing between the states $\left|-\frac{3}{2}\right\rangle$ and $\left|-\frac{1}{2}\right\rangle$ (GSLAC-1 and ESLAC-1), whereas the $\hat{V}_{\perp}^{(3)}$ term mixes the states $\left|-\frac{3}{2}\right\rangle$ and $\left|\frac{1}{2}\right\rangle$ (GSLAC-2 and ESLAC-2). The mixing matrix elements calculated in the first-order perturbation theory are equal to

$$\left\langle -\frac{3}{2}\left|\hat{V}_{\perp}^{(1)}\right|-\frac{1}{2}\right\rangle \approx \sqrt{3}D\theta \tag{11}$$

$$\left\langle -\frac{3}{2}\left|\hat{V}_{\perp}^{(2)}\right|-\frac{1}{2}\right\rangle \approx \sqrt{3}\kappa D\theta$$

$$\left\langle -3/2\left|\hat{V}_{\perp}^{(3)}\right|-1/2\right\rangle \approx -i\sqrt{3}\delta D\theta$$

Here $\kappa = \frac{g_\perp^{(2)}}{g_\parallel^{(1)}}$ and $\delta = \frac{g_\perp^{(3)}}{g_\parallel^{(1)}}$. These perturbations turn the GSLC-1,2 and ESLC-1,2 into the GSLAC-1,2 and ESLAC-1,2, respectively, see **Figures 2(A, B)**. In turn, these LACs give rise features in the field dependence of the PL intensity, see **Figure 2C**. The discussion of the precise choice of parameters, specifically $g$-factors, is presented in **Supplementary Materials**.

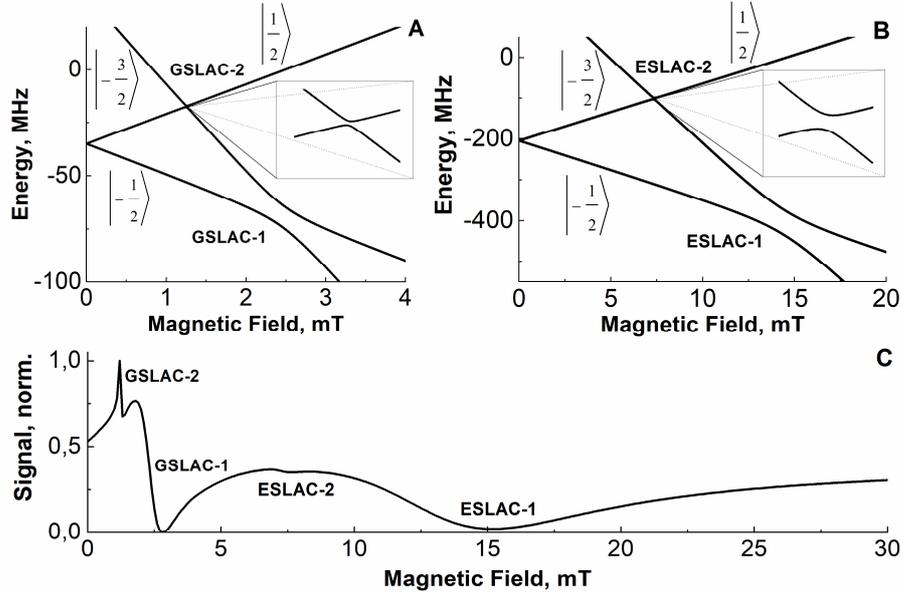

*Figure 2*. GSLACs (subplot A) and ESLACs (subplot B) in the $V_{Si}$ center. The LCs at $\gamma_e B = D_{E,G}$ and $\gamma_e B = 2D_{E,G}$ are turned into LACs by perturbation terms, in this example, by $\hat{V}_\perp = g_\perp^{(1)} \mu_B \hat{\mathbf{S}}_\perp \hat{\mathbf{B}}_\perp + g_\perp^{(2)} \mu_B \left\{ \hat{\mathbf{S}}_\perp \hat{\mathbf{B}}_\perp, \left( \hat{S}_z^2 - \frac{3}{4} \right) \right\} - \frac{i}{4} g_\perp^{(3)} \mu_B \left( \{ \hat{S}_+^2, \hat{S}_z \} B_+ - \{ \hat{S}_-^2, \hat{S}_z \} B_- \right)$. The LACs give rise to features in the field dependence of photoluminescence $I_{PL}(B)$, (normalized to 1), see subplot (C).

## III. Results and Discussion

In this section, we provide a comparison of theoretical results with available experimental data [43] on field-dependent PL and ODMR of the $V_{Si}$ centers.

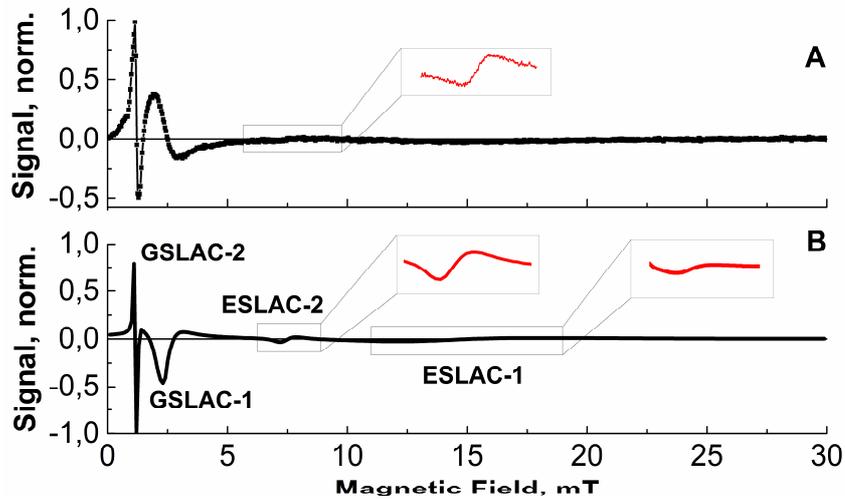

*Figure 3. Photoluminescence intensity $I_{PL}$ as a function of the static magnetic field B, applied parallel to the c-axis of the $V_{Si}$ center: experimental data (A) and simulation (B). Experimental data were recorded using lock-in detection; in the calculation we show the derivative of $I_{PL}$, i.e., $\frac{d}{dB}I_{PL}$, normalized to the maximum value. Features in the field dependence are assigned to the relevant LACs in GS and ES.*

The PL intensity plotted as a function of the external magnetic field is shown in **Figure 3**. To ease comparison with experimental data, we plot not the $I_{PL}(B)$ function, as shown in **Figure 2**, but the derivative $\frac{d}{dB}I_{PL}$. Under such conditions, each peak or dip in the field dependence is turned into a sharp feature with a positive component and a negative component (positive-negative feature for a peak and negative-positive feature for a dip). In the field dependence, we can clearly identify the lines coming from the GSLAC-1 and GSLAC-2, whereas the lines from the ESLAC-1 and ESLAC-2 are considerably weaker. Specifically, when the derivative $\frac{d}{dB}I_{PL}$ is plotted instead of $I_{PL}$, the smooth ESLAC features, in particular, the ESLAC-1 feature, become barely visible. The relative ratio of the GSLAC and ESLAC lines strongly depends on the relative values of the rates introduced in **Scheme 1**. The type of the feature, i.e., peak or dip, depend on the relation between $k_1^{fl}$ and $k_2^{fl}$ and also between $k_1^{isc}$ and $k_2^{isc}$ (the features can only show up when $k_1^{fl} \neq k_2^{fl}$ or $k_1^{isc} \neq k_2^{fl}$). The calculated field dependence is in good agreement with the experimental curve: the positions of the features are properly reproduced, as well as their relative intensities and widths.

In addition to the field dependence of PL, we performed calculation of ODMR-related phenomena.

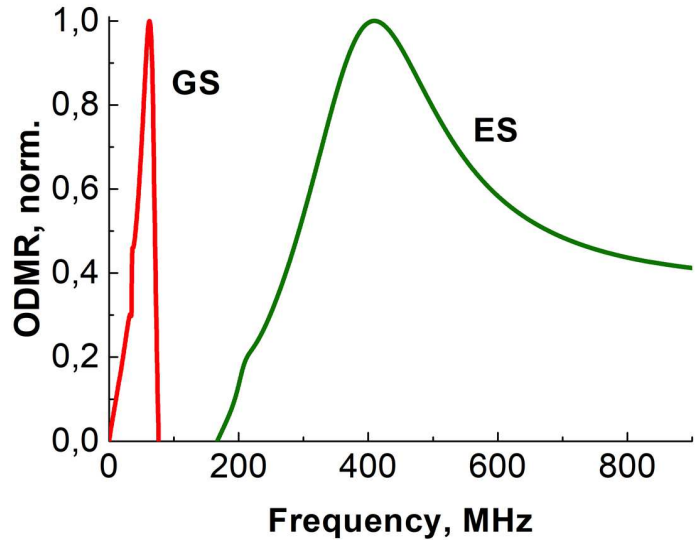

*Figure 4. ODMR spectrum, calculated for zero external magnetic field; the two components correspond to transitions between the ZFS states in GS and ES. The ODMR intensity is normalized to 1 for each transition.*

First, we consider the case where the external static magnetic field is zero. In this situation, by using the method outlined above, we calculated ODMR spectra, see **Figure 4**, with the same parameters as in **Figure 3**. In the ODMR spectrum obtained at zero field there are two pronounced lines, corresponding to the transitions between the ZFS states with different absolute values of $S_z$, i.e., with $|S_z| = \frac{1}{2}$ and $|S_z| = \frac{3}{2}$. In both ES and GS, these states are populated differently, owing to the properties of the optical cycle. The RF-driven transitions between these states alter the state populations, affecting the PL intensity. In the spectrum, see **Figure 4**, we can see two lines at $\omega_{rf}/2\pi \approx 70$ MHz and $\omega_{rf}/2\pi \approx 410$ MHz, corresponding to the RF-driven transitions between the ZFS states in GS and ES, respectively. The feature at 70 MHz is significantly narrower than that at 410 MHz. Such an appearance of the ODMR spectrum is in accordance with experimental spectra [43], which are not shown here.

Second, we calculated the ODMR spectrum under the conditions where $\omega_{rf}$ is fixed, but the magnetic field is swept. The ODMR spectrum is then given by the relative ODMR intensity, i.e., by the variation of the PL, obtained upon application of the external RF-field, which is plotted against the magnetic field strength $B$. Such a curve in shown in **Figure 5**, along with the experimental dependence. In this curve, four features emerge, which correspond to the four relevant LACs, GSLAC-1,2 and ESLAC-1,2. Two features are narrow, namely, GSLAC-2 and ESLAC-2, and two features are broad, namely, GSLAC-1 and ESLAC-1. The calculated curve is in good agreement with experimental data [43]. In experimental and theoretical ODMR spectra one can clearly see the features originating from GSLAC-1 and ESLAC-1 (the former is narrower and the latter is broader), while the features associated with GSLAC-2 and ESLAC-2 are barely visible.

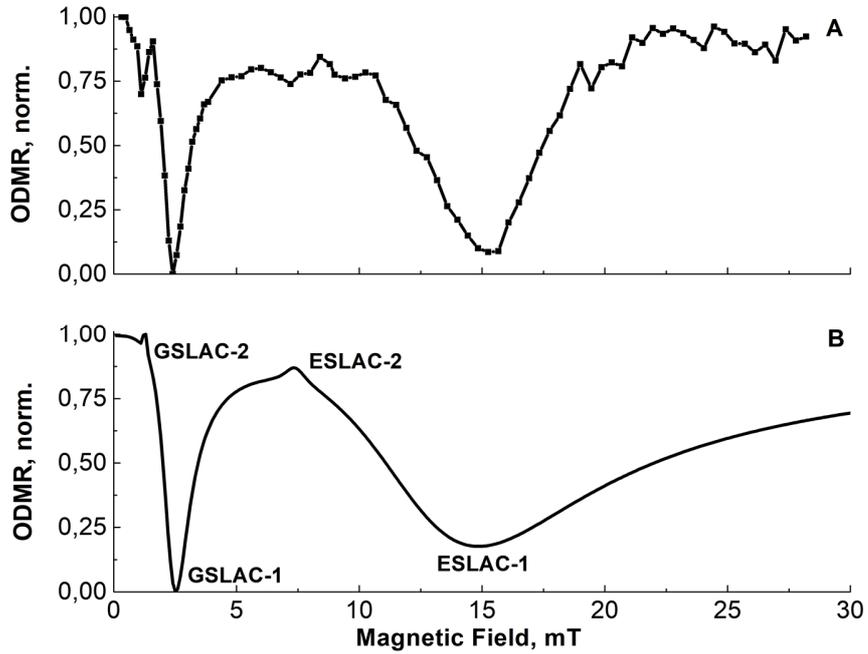

*Figure 5. Normalized relative ODMR signal as a function of the static magnetic field B applied parallel to the c axis of the $V_{Si}$ center: experimental data (A) and simulation (B). Features in the field dependence are assigned to the relevant LACs in GS and ES.*

## IV. Summary and Conclusions

In this work, we present a theoretical formalism, which is capable of describing spin-dependent phenomena in color centers in SiC. Here we focus on the silicon-vacancy centers, $V_{Si}$ or V2 centers, having spin-$\frac{3}{2}$ ground state and excited state, which can be polarized by light excitation. Our theoretical method explicitly treats the spin dynamics in both states as well as spin-dependent transitions between the GS, ES and MS states: this is done by using Lindblad-type super-operators. We demonstrate that the proposed approach can be used to simulate a number of spin-dependent phenomena, namely, magnetic field dependent PL intensity and ODMR signals. In the field dependence of the PL intensity sharp peaks and dips are present, which can be assigned to LACs in ES and GS. We calculate zero-field ODMR spectra obtained by sweeping the frequency of the applied RF-field and also relative changes in the ODMR signals, obtained upon variation of the external magnetic field. In the latter case, features associated with the GSLACs and ESLACs become visible again. All calculation results are in good agreement with previously reported experimental data [43]. The precise shape of LAC-derived features as well as their phase (peak or dip) depends on the rates of the transitions between GS, ES and MS.

We anticipate that the proposed method can be utilized to describe spin-dependent phenomena in other defect color centers. One more application of the present theoretical approach is the analysis of light-induced nuclear spin polarization: such polarization can be observed, for instance, in diamond crystals containing NV⁻ centers and utilized for dramatic enhancement of weak NMR signals [12-17].

### Acknowledgements

This work has been supported by the Russian Science Foundation (grant No. 20-63-46034). We acknowledge Prof. Sergey Tarasenko for stimulating discussions and Prof. V. Dyakonov and Dr. G. Astakhov for providing experimental data on PL and ODMR.

**Supplementary Materials** for the article

# Level-crossing induced spin phenomena in SiC: a theoretical study

Denis V. Sosnovsky, Konstantin L. Ivanov*

*International Tomography Center, Siberian Branch of the Russian Academy of Sciences, Novosibirsk, 630090, Russia*

*Novosibirsk State University, Novosibirsk, 630090, Russia*

**Contents**

## A. Lindblad super-operators

The Lindblad operator $\hat{\mathcal{L}}$ is defined in the basis of states $|\mu_{GS}\rangle$, $|\mu_{ES}\rangle$, $|\mu_{MS}\rangle$. Since the rates of the transitions do not depend on the nuclear spin state, the operator $\hat{\mathcal{L}}$ is given by the Kronecker (direct) product of the electronic Lindblad operator $\hat{\mathcal{L}}_{el}$ (9×9 matrix) and unity matrix corresponding to the nuclear spin subsystem, here by the 2×2 unity matrix (as we deal with nuclear spin $\frac{1}{2}$).

The operator $\hat{\mathcal{L}}_{el}$ is defined in the basis of electronic spin states

$$\left|\tfrac{3}{2}\right\rangle_{GS}, \left|\tfrac{1}{2}\right\rangle_{GS}, \left|-\tfrac{1}{2}\right\rangle_{GS}, \left|-\tfrac{3}{2}\right\rangle_{GS}, \left|\tfrac{3}{2}\right\rangle_{ES}, \left|\tfrac{1}{2}\right\rangle_{ES}, \left|-\tfrac{1}{2}\right\rangle_{ES}, \left|-\tfrac{3}{2}\right\rangle_{ES}, |MS\rangle \tag{S1}$$

To introduce the explicit for of this operator, we introduce the rates, $k_{mn}$, of the transitions between the electronic quantum states $|m\rangle_{el} \to |n\rangle_{el}$. After that, we specify the corresponding non-zero elements of the Lindblad operator are $\langle n|\hat{\mathcal{L}}_{el}|m\rangle = \sqrt{k_{mn}}$. As a result, we obtain the following matrix:

$$\hat{\mathcal{L}}_{el} = \begin{pmatrix} 0 & 0 & 0 & 0 & \sqrt{k_2^{fl}} & 0 & 0 & 0 & 0 \\ 0 & 0 & 0 & 0 & 0 & \sqrt{k_1^{fl}} & 0 & 0 & \sqrt{k_{isc}'} \\ 0 & 0 & 0 & 0 & 0 & 0 & \sqrt{k_1^{fl}} & 0 & 0 \\ 0 & 0 & 0 & 0 & 0 & 0 & 0 & \sqrt{k_2^{fl}} & \sqrt{k_{isc}'} \\ \sqrt{I} & 0 & 0 & 0 & 0 & 0 & 0 & 0 & 0 \\ 0 & \sqrt{I} & 0 & 0 & 0 & 0 & 0 & 0 & 0 \\ 0 & 0 & \sqrt{I} & 0 & 0 & 0 & 0 & 0 & 0 \\ 0 & 0 & 0 & \sqrt{I} & 0 & 0 & 0 & 0 & 0 \\ 0 & 0 & 0 & 0 & \sqrt{k_1^{isc}} & \sqrt{k_2^{isc}} & \sqrt{k_2^{isc}} & \sqrt{k_1^{isc}} & 0 \end{pmatrix} \tag{S2}$$

This form of the operator has been used for solving the equation for the density matrix.

## B. LACs in the presence of hyperfine couplings

In this subsection, we analyze how different interactions, notably, the hyperfine coupling (HFC), affect the relevant LACs and the spin dynamics in the $V_{Si}$ center.

Assuming that the $Z$-axis of Zero-Field Splitting (ZFS) tensor is perfectly aligned with the $Z$-axis of laboratory frame $(Z||\boldsymbol{B})$ and taking into account only secular HFC, for a $V_{Si}$ center with a single spin-$\frac{1}{2}$ nucleus we obtain the following energies of the spin states for GS and ES:

$$
\begin{aligned}
E_1 &= D + \frac{3}{2}\omega_e - \frac{1}{2}\omega_N + \frac{3}{4}A_{zz}, & E_2 &= D + \frac{3}{2}\omega_e + \frac{1}{2}\omega_N - \frac{3}{4}A_{zz}, \\
E_3 &= -D + \frac{1}{2}\omega_e - \frac{1}{2}\omega_N + \frac{1}{4}A_{zz}, & E_4 &= -D + \frac{1}{2}\omega_e + \frac{1}{2}\omega_N - \frac{1}{4}A_{zz}, \\
E_5 &= -D - \frac{1}{2}\omega_e - \frac{1}{2}\omega_N - \frac{1}{4}A_{zz}, & E_6 &= -D - \frac{1}{2}\omega_e + \frac{1}{2}\omega_N + \frac{1}{4}A_{zz}, \\
E_7 &= D - \frac{3}{2}\omega_e - \frac{1}{2}\omega_N - \frac{3}{4}A_{zz}, & E_8 &= D - \frac{3}{2}\omega_e + \frac{1}{2}\omega_N + \frac{3}{4}A_{zz}.
\end{aligned}
\tag{S3}
$$

Here $A_{zz}$ is the secular HFC term, $\omega_e = \gamma_e B$ and $\omega_N = \gamma_N B$ stand for the Larmor precession frequencies of the free electron and the nucleus. The notation of the states in eq. (S1) is as follows. State $|1\rangle$ corresponds to $\left|\frac{3}{2}, \alpha\right\rangle$ at high magnetic fields (i.e., in the limit $\gamma_e B \gg D$); state $|2\rangle$ goes to $\left|\frac{3}{2}, \beta\right\rangle$ at high fields; for the other states the correlation with high field states is as follows: $|3\rangle \to \left|\frac{1}{2}, \alpha\right\rangle$, $|4\rangle \to \left|\frac{1}{2}, \beta\right\rangle$, $|5\rangle \to \left|-\frac{1}{2}, \alpha\right\rangle$, $|6\rangle \to \left|-\frac{1}{2}, \beta\right\rangle$, $|7\rangle \to \left|-\frac{3}{2}, \alpha\right\rangle$, $|8\rangle \to \left|-\frac{3}{2}, \beta\right\rangle$.

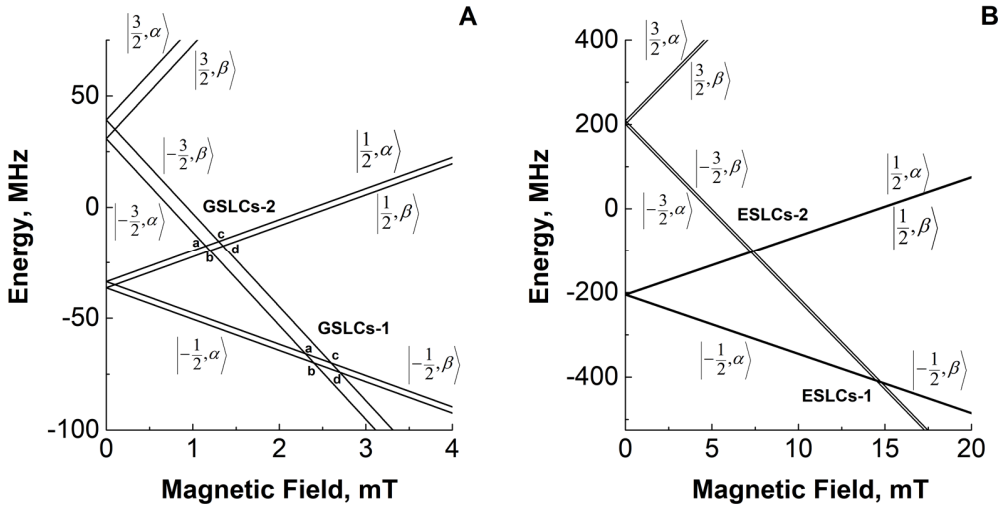

**Figure S1**. Energy levels and LCs in the $V_{Si}$ center shown for GS (A) and ES (B). Here $A_{zz} = 0.2$ mT.

When nuclear spin states are taken into account, the two Level Crossings (LCs) of interest split into four LCs. Hence, instead of the single LC-1 and LC-2 we obtain two groups of LCs; the four LCs points are the vertices of parallelograms, see **Figure S1 (A)**. The relevant LCs-1 in the system are the following ones: LC-1a of $E_6$ and $E_7$, LC-1b of $E_5$ and $E_7$, LC-1c of $E_6$ and $E_8$, LC-1d of $E_5$ and $E_8$. The relevant LC2s in the system are introduced in a similar way: LC-2a of $E_3$ and $E_7$, LC-2b of $E_4$ and $E_7$, LC-2c of $E_3$ and $E_8$, LC-2d of $E_4$ and $E_8$. To determine the positions of all four LCs we need to solve linear equations, which give rise to the following results:

$$B_{LC-1}^{(a)} = \frac{2D-a}{\gamma_e+\gamma_n}, \quad B_{LC-1}^{(b)} = \frac{2D-\frac{a}{2}}{\gamma_e}, \quad B_{LC-1}^{(c)} = \frac{2D+\frac{a}{2}}{\gamma_e}, \quad B_{LC-1}^{(d)} = \frac{2D+a}{\gamma_e-\gamma_n} \quad \text{(S4)}$$

$$B_{LC-2}^{(a)} = \frac{D-\frac{a}{2}}{\gamma_e}, \quad B_{LC-2}^{(b)} = \frac{D-\frac{a}{4}}{\gamma_e+\frac{\gamma_n}{2}}, \quad B_{LC-2}^{(c)} = \frac{D+\frac{a}{4}}{\gamma_e-\frac{\gamma_n}{2}}, \quad B_{LC-2}^{(d)} = \frac{D+\frac{a}{2}}{\gamma_e};$$

Various perturbations cause mixing of the spin states and turn the eight relevant LCs into LACs. One should note that mixing at four of the eight LACs, namely, LAC-1a and LAC-1d originating from LC-1a and LC-1d, and LAC-2b and LAC-2c originating from LC-2b and LC-2c, lead to nuclear spin flips causing polarization transfer from the electron subsystem to the nuclear spins, which might be of interest for nuclear spin hyperpolarization experiments. Below we consider how various perturbations which affect energy levels. In each case we also calculate the matrix elements responsible for spin mixing and model the field dependence of the photoluminescence intensity of the color center.

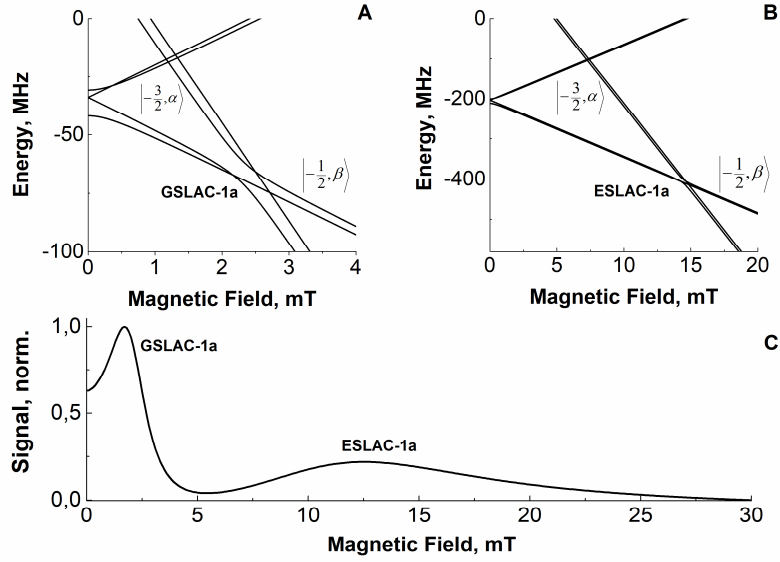

*Figure S2*. GSLACs (subplot A) and ESLACs (subplot B) in the V$_{Si}$ center. Here LCs-1a at $(\gamma_e+\gamma_n)B = 2D_{E,G} - a$ are turned into LACs-1a (in GS and ES) by the perturbation term $\frac{1}{2}(\hat{S}_+\hat{I}_- + \hat{S}_-\hat{I}_+)$. The LACs give rise to two features in the field dependence of photoluminescence $I_{PL}(B)$, (normalized to unity), see subplot C. Here $A_{xx} = A_{yy} = A_{zz} = 0.2\ mT$.

### (i) Case B||Z, isotropic HFC

In the simplest case we take into account only isotropic HFC, i.e., in the $\hat{\mathcal{A}}$ tensor the diagonal elements are identical, $A_{xx} = A_{yy} = A_{zz} = A_{iso}$, and all other elements are zero. In this situation, only electron-nuclear flip-flops are possible because isotropic HFC contains only terms $\frac{1}{2}(\hat{S}_+\hat{I}_- + \hat{S}_-\hat{I}_+)$. These terms give rise to mixing of the $\left|-\frac{3}{2},\alpha\right\rangle$ and $\left|-\frac{1}{2},\beta\right\rangle$ states as the z-projection of the total spin, $\hat{\mathbf{F}} = \hat{\mathbf{S}} + \hat{\mathbf{I}}$, of the system is conserved. Hence, only one of the eight LCs, namely LC-1a, is turned into a LAC, see **Figure S2 (A, B)**. As a consequence, in the PL field dependence there are two peaks corresponding to the LAC-1a in the ground and excited electronic states, see **Figure S2 (C)**. The width of the feature is given by the width of the LAC region, i.e., by the $A_{iso}$ value. The mixing matrix element between the relevant states is equal to

S3

$$V_{iso} = \frac{\sqrt{3}}{4}(A_{xx} + A_{yy}) = \frac{\sqrt{3}}{2} A_{iso} \qquad (S5)$$

**(ii) Case B||Z, anisotropic HFC**

Now let us turn to cases of higher complexity, where other perturbation terms are present in the Hamiltonian. When we assume anisotropic HFC so that $A_{xx} \neq A_{yy}$, additional HFC terms emerge, namely the $\hat{S}_+\hat{I}_+$ and $\hat{S}_-\hat{I}_-$ terms. Such terms can mix the $|-3/2,\beta\rangle$ and $|-1/2,\alpha\rangle$ states. As a consequence, one more LC turns into an LAC, see **Figure S3 (A, B)**. The corresponding mixing matrix element is equal to

$$\frac{A_{xx} - A_{yy}}{4} \langle -3/2, \beta | \hat{S}_+\hat{I}_+ + \hat{S}_-\hat{I}_- | -1/2, \alpha \rangle = \frac{\sqrt{3}}{4}(A_{xx} - A_{yy}) \qquad (S6)$$

The calculated PL field dependence shown in **Figure S3 C** is in accordance with this expectation: there is a structure originating from LAC-1a LAC-1d for the excited and ground state of the SIC center. Different widths of the features are due to the different mixing matrix elements; the mixing term is greater for LAC-1a for the chosen set of parameters.

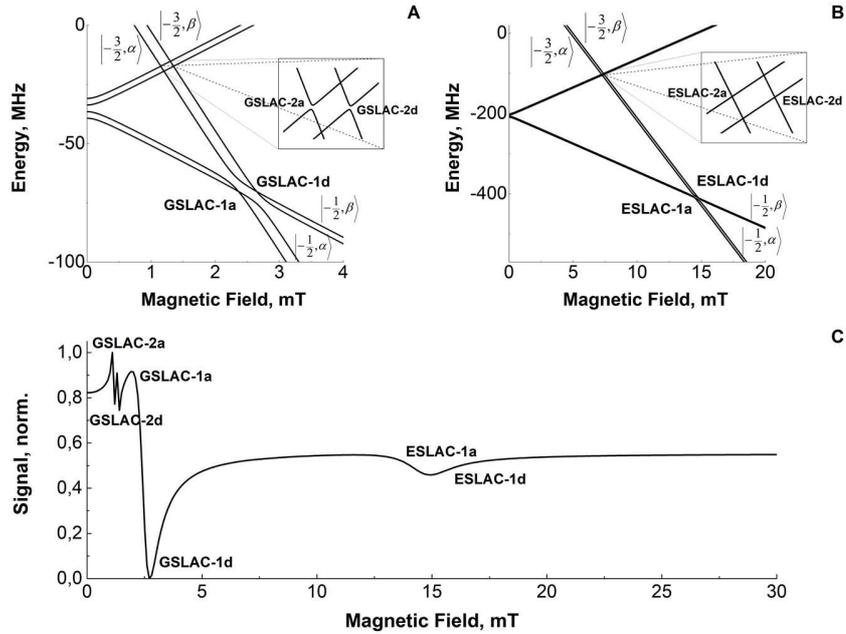

*Figure S3*. GSLACs (subplot A) and ESLACs (subplot B) in the $V_{Si}$ center. The LCs-1a at $(\gamma_e + \gamma_n)B = 2D_{E,G} - a$ are turned into LACs-1a (in GS and ES) by the perturbation term $\frac{1}{2}(\hat{S}_+\hat{I}_- + \hat{S}_-\hat{I}_+)$, the LCs-1d at $(\gamma_e - \gamma_n)B = 2D_{E,G} + a$ are turned into LACs-1d by the perturbation terms $\hat{S}_+\hat{I}_+$ and $\hat{S}_-\hat{I}_-$. The LCs-2a at $\gamma_e B = D_{E,G} - a/2$ are turned into LACs-2a, the LCs-2d at $\gamma_e B = D_{E,G} + a/2$ are turned into LACs-2d. The LACs give rise to features in the field dependence of photoluminescence $I_{PL}(B)$, (normalized to unity), see subplot C. Here $A_{xx} = A_{zz} = 0.2$ mT, $A_{yy} = 0$.

Mixing between the states $|-3/2,\alpha\rangle$ and $|1/2,\alpha\rangle$ at $B_{LC-2}^{(a)}$ and between the states $|-3/2,\beta\rangle$ and $|1/2,\beta\rangle$ at $B_{LC-2}^{(d)}$ also gives rise to variation of the photoluminescence intensity. At a first glance, $\hat{S}_+\hat{I}_+$ and $\hat{S}_-\hat{I}_-$ terms cannot mix the states of interest. The reason is that these terms flip electron spin changing the $S_z$ value such that $\Delta S_z = 1$, but not $\Delta S_z = 2$. However, a more detailed analysis in the framework of the perturbation theory shows that these terms also generate mixing of the degenerate states at LC-2a and LC-

2d. The reason is that these terms first mix the following states $|1/2, \alpha\rangle \leftrightarrow |-1/2, \beta\rangle, |-3/2, \beta\rangle \leftrightarrow |-1/2, \alpha\rangle$. Consequently, the eigenstates are modified and then the new perturbed states get mixed by the isotropic HFC term. Effectively this leads to non-zero matrix elements: $\langle -3/2, \alpha|\widehat{\mathcal{H}}|1/2, \alpha\rangle \neq 0$, $\langle -3/2, \beta|\widehat{\mathcal{H}}|1/2, \beta\rangle \neq 0$. Consequently, the combination of isotropic and anisotropic HFC turns LC-2a and LC-2d into LAC-2a and LAC-2d, respectively, see **Figure S3 (A, B)**. The LAC-2s give rise to features in the photoluminescence intensity. Specifically, one can see two new features in the field dependence of $I_{PL}$ (clearly visible only for the ground state of the V$_{Si}$ center due to the small values of HFC, as compared to the ZFS parameters in the excited state), see **Figure S3 C**.

Detailed calculation of the coupling matrix elements is given below. The HFC terms have the following non-zero matrix elements in the initial basis of spin states:

$$V_1 = \frac{A_{xx} + A_{yy}}{4} \left\langle -\frac{3}{2}, \alpha \middle| \hat{S}_- \hat{I}_+ \middle| -\frac{1}{2}, \beta \right\rangle = \frac{\sqrt{3}}{4} (A_{xx} + A_{yy}),$$

$$V_2 = \frac{A_{xx} - A_{yy}}{4} \left\langle -\frac{3}{2}, \beta \middle| \hat{S}_- \hat{I}_- \middle| -\frac{1}{2}, \alpha \right\rangle = \frac{\sqrt{3}}{4} (A_{xx} - A_{yy}),$$

$$V_3 = \frac{A_{xx} + A_{yy}}{4} \left\langle \frac{1}{2}, \alpha \middle| \hat{S}_- \hat{I}_+ \middle| \frac{3}{2}, \beta \right\rangle = \frac{\sqrt{3}}{4} (A_{xx} + A_{yy}),$$

$$V_4 = \frac{A_{xx} - A_{yy}}{4} \left\langle \frac{1}{2}, \alpha \middle| \hat{S}_+ \hat{I}_+ \middle| -\frac{1}{2}, \beta \right\rangle = \frac{1}{2} (A_{xx} - A_{yy}),$$

$$V_5 = \frac{A_{xx} + A_{yy}}{4} \left\langle \frac{1}{2}, \beta \middle| \hat{S}_+ \hat{I}_- \middle| -\frac{1}{2}, \alpha \right\rangle = \frac{1}{2} (A_{xx} + A_{yy}),$$

$$V_6 = \frac{A_{xx} - A_{yy}}{4} \left\langle \frac{1}{2}, \beta \middle| \hat{S}_- \hat{I}_- \middle| \frac{3}{2}, \alpha \right\rangle = \frac{\sqrt{3}}{4} (A_{xx} - A_{yy}).$$

(S7)

Hence, the HFC terms perturb the initial spin states. The true eigen-states are then as follows:

$$|-3/2, \alpha\rangle' \approx |-3/2, \alpha\rangle + q_1|-1/2, \beta\rangle,$$

$$|-3/2, \beta\rangle' \approx |-3/2, \beta\rangle + q_2|-1/2, \alpha\rangle,$$

$$|1/2, \alpha\rangle' \approx |1/2, \alpha\rangle + q_3|3/2, \beta\rangle + q_4|-1/2, \beta\rangle,$$

$$|1/2, \beta\rangle' \approx |1/2, \beta\rangle + q_5|-1/2, \alpha\rangle + q_6|3/2, \alpha\rangle.$$

(S8)

Here (as follows from the perturbation theory)

$$q_1 = \frac{V_1}{\omega_e - \omega_n}, q_2 = \frac{V_2}{\omega_e + \omega_n}, q_3 = -\frac{V_3}{2D + \omega_e + \omega_N - A_{zz}},$$

$$q_4 = \frac{V_4}{\omega_e - \omega_n}, q_5 = \frac{V_5}{\omega_e + \omega_n}, q_6 = -\frac{V_6}{2D + \omega_e - \omega_n + A_{zz}}$$

(S9)

where the matching conditions for LC2s were used: $2\omega_e = 2D - A_{zz}$ (for $q_1$ and $q_4$), $2\omega_e = 2D + A_{zz}$ (for $q_2$ and $q_5$). In the new basis the Hamiltonian has the following non-zero matrix elements:

$$\left\langle -\frac{3}{2},\alpha \middle| \widehat{\mathcal{H}} \middle| \frac{1}{2},\alpha \right\rangle = q_1^* V_4 + q_4 V_1 + q_1^* q_4 E_{-\frac{1}{2},\beta} \tag{S10}$$

$$\langle -3/2,\beta | \widehat{\mathcal{H}} | 1/2,\beta \rangle = q_2^* V_5 + q_5 V_2 + q_2^* q_5 E_{-1/2,\alpha}.$$

Neglecting the small $\omega_N$ value (as compared to the $D$ values and HFC terms) we can obtain the following approximate expressions for the matrix elements of interest:

$$\langle -3/2,\alpha | \widehat{\mathcal{H}} | 1/2,\alpha \rangle = \frac{\sqrt{3}}{4} \frac{(D - A_{zz})(A_{xx}^2 - A_{yy}^2)}{(2D - A_{zz})^2}, \tag{S11}$$

$$\langle -3/2,\beta | \widehat{\mathcal{H}} | 1/2,\beta \rangle = \frac{\sqrt{3}}{4} \frac{(D + A_{zz})(A_{xx}^2 - A_{yy}^2)}{(2D + A_{zz})^2}$$

A similar analysis can be done for other terms of the anisotropic HFC tensor, for instance for $A_{zx}\hat{S}_z\hat{I}_x$ term [S1].

## C. Calculation parameters

For the electronic Zeeman interaction, we used the following parameters:

$$g_{\parallel}^{(1)} = 2.0, \quad g_{\parallel}^{(2)} = 0.0, \quad g_{\parallel}^{(3)} = 0.0$$
$$g_{\perp}^{(1)} = 2.0, \quad g_{\perp}^{(2)} = 0.0, \quad g_{\perp}^{(3)} = 0.2 \tag{S12}$$

These values are in accordance with literature data [S2]. All g-factors except $g_{\parallel}^{(3)}$ were taken from Ref. [S2]. In the original work $g_{\parallel}^{(3)} = 0.6$, but in the calculations this parameter was taken zero to achieve the best agreement with the experimental data.

One should note that the term $\hat{V}_{\parallel}^{(3)} = g_{\parallel}^{(3)} \mu_B \frac{\hat{S}_+^3 - \hat{S}_-^3}{4i} B$ causes mixing of only the states $\left|\frac{3}{2}\right\rangle$ and $\left|-\frac{3}{2}\right\rangle$ and due to the small value of $g_{\parallel}^{(3)}$ slightly shifts the position of the relevant LCs. In our calculation in fact we neglect the small effect of this perturbation term.

The ZFS parameters $D_G = 1.25$ mT and $D_E = 7.32$ mT are in accordance with Ref. [S3].

Considering the effect of misalignment of the crystal we used the tilt angle $\theta = 5^0$. Here $\theta$ is the angle between the $Z$-axis of the ZFS tensor and the external magnetic field (we assume that the **B** vector lies in the $XZ$-plane); this angle is taken small, hence $\sin \theta \approx \theta$. The $\theta = 5^0$ value was chosen to give the best agreement with the experimental data.

For the HFC terms we use the following values

$$A_{xx}^{(GS)} = A_{yy}^{(GS)} = A_{zz}^{(GS)} = 0.001 \text{ mT}, \quad A_{xx}^{(ES)} = A_{yy}^{(ES)} = A_{zz}^{(ES)} = 0.001 \text{ mT} \tag{S13}$$

The reason is that natural abundance of the magnetic silicone isotope, $^{29}$Si, is only 4.7%; in some cases, 4H-SiC crystals are studied for 99.0% of the $^{28}$Si isotope with zero nuclear spin. Hence, the concentration of magnetic nuclei is very low, and the effective HFC terms are also low. Hence, in calculation the isotropic HFC was chosen very small (but not exactly zero). Pseudo-secular couplings were taken zero for the sake of simplicity. In Section B of this file, for generality, we also consider the role of different interaction assuming higher HFC terms.

For the transition rates we used the following parameters, here given in units of ns$^{-1}$:

$$I = 0.01, \quad k_2^{fl} = 2k_1^{fl} = 0.1, \quad k_1^{isc} = 20k_2^{isc} = 0.2, \quad k'_{isc} = 0.01 \tag{S14}$$

The reaction rates were chosen to give the best agreement with the experimental data. The order of the rate constants is in accordance with literature data [S4].